\documentclass[fleqn,12pt]{wlscirep}
\usepackage[utf8]{inputenc}
\usepackage[T1]{fontenc}
\usepackage{caption}
\usepackage{subcaption}
\usepackage{graphicx}
\usepackage{dcolumn}
\usepackage{bm}
\usepackage{color}
\usepackage{amsmath}
\usepackage{newtxtext,newtxmath} 
\usepackage{algorithm}
\usepackage{algpseudocode}
\usepackage{adjustbox}
\usepackage{hyperref}
\DeclareMathOperator*{\argmax}{argmax}

\newcommand{\blue}[1]{\textcolor{blue}{#1}}
\newcommand{\mg}[1]{{#1}}

\title{Depolarization of echo chambers by random dynamical nudge}

\author[1,2,3]{Christopher Brian Currin}
\author[4,5]{Sebastián Vallejo Vera}
\author[6,*]{Ali Khaledi-Nasab}

\affil[1]{Department of Human Biology, Faculty of Health Sciences, University of Cape Town, South Africa}
\affil[2]{Neuroscience Institute, University of Cape Town, South Africa}
\affil[3]{Institute of Science and Technology Austria, Klosterneuburg, Lower Austria, Austria}
\affil[4]{Tecnologico de Monterrey, School of Social Science and Government}
\affil[5]{Interdisciplinary Laboratory of Computational Social Science, Mexico}
\affil[6]{Stanford University, Stanford, CA}

\affil[*]{ali.khaledi1989@gmail.com}

\keywords{opinion dynamics, Nudge, echo chambers}

\begin{abstract}

In social networks, users often engage with like-minded peers. This selective exposure to opinions might result in echo chambers, i.e., political fragmentation and social polarization of user interactions. When echo chambers form, opinions have a bimodal distribution with two peaks on opposite sides. In certain issues, where either extreme positions contain a degree of misinformation, neutral consensus is preferable for promoting discourse. In this paper, we use an opinion dynamics model that naturally forms echo chambers in order to find a feedback mechanism that bridges these communities and leads to a neutral consensus. \mg{We introduce \textit{random dynamical nudge} (RDN), which presents each agent with input from a random selection of other agents' opinions and does not require surveillance of every person's opinions.} Our computational results \mg{in two different models} suggest that the RDN leads to a unimodal distribution of opinions centered around the neutral consensus. Furthermore, the RDN is effective both for preventing the formation of echo chambers and also for depolarizing existing echo chambers. Due to the simple and robust nature of the RDN, social media networks might be able to implement a version of this self-feedback mechanism, when appropriate, to prevent the segregation of online communities on complex social issues.

\end{abstract}
\begin{document}

\flushbottom
\maketitle
\thispagestyle{empty}

\noindent Keywords: Social networks, depolarization, echo chambers, nudge

\section*{Introduction}

Online social media has transformed the way we communicate and consume information. Despite the potential to democratize information sharing and inform public opinion, users often are selectively exposed to cognitively congruent content to their own views \cite{himelboim2013tweeting, aruguete2018time}. This tendency reduces content diversity \cite{cinelli2020covid} and can lead to polarization\cite{perra2019modelling,sasahara2019inevitability}, creating clusters of like-minded individuals known as echo chambers \cite{jamieson2008echo, garrett2009echo}. Polarization is exacerbated during times of concentrated attention (e.g., debates over controversial issues, political events) \cite{cardenal2019echo}, which creates space for the spread of misinformation \cite{bessi2016homophily} and potentially hampers the democratic deliberative process \cite{sunstein2007republic,cinelli2020echo}.

Even though echo chambers are not a permanent fixture of social media \cite{guess2021almost,del2016spreading,barbera2015tweeting}, studies have shown that the formation of echo chambers have pernicious effects, fostering the propagation of propaganda and ``\textit{fake news},'' \cite{wang2020viral} increasing gender asymmetries \cite{usher2018twitter}, and polarizing political camps \cite{cinelli2021echo,du2016echo,balsamo2019inside}. The latter is intensified when the nature of the issue discussed politically divides the population (e.g., presidential elections, gun control, health care)\cite{barbera2015tweeting,garimella2018political}. Furthermore, the spread of misinformation can have real-life consequences when individuals act upon it\cite{tuters2018post}. 

Alternatively, clustering around a shared central space may decrease the distance between users holding opposing opinions, decrease the propagation speed of harmfully false information, while also increasing the dialogue between communities informed by less pernicious content. In certain issues, the neutral consensus, or the middle point, is favorable. Since the formation of echo chambers in social media networks implies a lack of communication across communities\cite{cinelli2020echo, cinelli2021echo}, we propose a mechanism to avoid the formation of echo chambers - and to disband those already formed - by presenting each agent with input from a random selection of other agents' opinions. \mg{This mechanism, which we term \textit{random dynamical nudge} (RDN, $\mathcal{R}$), aims to push a system of polarized opinions with few interactions between users at different ends of the distribution (i.e., bimodal distribution) towards a normal distribution with most of the interactions taking place close to a shared neutral consensus. The ultimate goal of the RDN is not to homogenize divergent opinions but rather to \textit{increase} dialogue across different positions based on content that might be ideologically slanted yet devoid of misinformation.} The potential outcomes of the RDN are democratic deliberation, consensus building, and the avoidance of the negative effects from unwavering, polarized users.

\mg{The main idea of the paper and the motivation for the RDN come from research in behavioral economics showing how subtle nudges can significantly affect individuals' behavior, even when they know that they are receiving a random nudge \cite{thaler2009nudge,kahneman2011thinking}. Formally, a nudge is any form of choice architecture that alters people's behavior in a predictable way without restricting options or significantly changing their economic incentives. It is shown that judgment of individuals is heavily affected by such cues \cite{thaler2009nudge}.} One example is the \textit{anchoring effect}, where a random input skews the individuals' judgments toward themselves \cite{kahneman2011thinking,thaler2009nudge, tversky1974judgment,lorenz2020behavioural}. Integrating epistemic cues -- an educative nudge -- can lead to more informed choices \cite{lorenz2020behavioural}. The extant literature indicates that inter-group contact can increase deliberation and compromise \cite{gronlund2015does,huckfeldt2004political}, as well as challenge stereotypes that develop in the absence of positive interactions among opposing groups \cite{pettigrew2006meta}. Studies reveal that users are more likely to polarize when they receive feedback based on their ideological views \cite{bakshy2015exposure}. However, users are not actively avoiding challenges to their own opinion nor are they necessarily polarized by opposing views\cite{garrett2009echo}. Thus, to depolarize networks, one needs to encourage dialogue (i.e., challenges to the agent's views) and interaction across communities. \mg{In a system where the opinions are segregated, providing a random sample of opinions from other users in the network might be helpful in depolarizing the social networks.}

In this paper, we present an extension of a recently introduced opinion dynamics model, and we numerically explore the possible effects of random nudge interventions. Baumann et al. \cite{baumann2020modeling,baumann2021emergence} proposed a simple opinion dynamics model which they validated against Twitter's data. Here we numerically explore possible effects of opinion-randomization interventions. Baumann et al.'s \cite{baumann2020modeling,baumann2021emergence} model gives rise to echo chambers from interactions among individuals. This model does not assume a structure between the correlated activity among the agents in our network. This framework has been used successfully not only to replicate echo chambers in Twitter \cite{baumann2020modeling} but also to study the emergence of issue alignment \cite{baumann2021emergence}. In this model, opinions evolve according to the interactions among agents, and the evolution is mediated by the degree of homophily, i.e., two agents with similar opinions have a higher chance of interacting \cite{holme2006nonequilibrium,kimura2008coevolutionary}. The evolution of opinions based on social interaction leads to a feedback loop which in turn leads to a correlation between the distribution of opinions and the network structure \cite{baumann2020modeling,baumann2021emergence}. Moreover, following the group theory of polarization\cite{isenberg1986group,myers1976group} interacting agents sharing similar views can reinforce the mutual stance \cite{vinokur1974effects}. Opinion dynamic models such as this one have been widely used to simulate the behavior of agents in public debates \cite{galam2005local,galam2016stubbornness,castellano2009statistical,martins2013building}, including the polarization of public opinion \cite{galam2005heterogeneous}. We take advantage of an opinion dynamics model that accurately simulates the formation of echo chambers in social media environments\cite{baumann2020modeling} and propose an intervention (i.e., RDN) to counter this phenomenon.

\mg{As we point out in the paper, one particularly relevant characteristic of our RDN is its feasibility as an applied tool to be used by social media platforms when addressing polarized communities and misinformation. The intervention relies on information that comes from within the network, and its application requires little supervision. In fact, RDN exploits the effectiveness of nudges to modify behavior, and randomizes the exposure to opinions, a process that can be easily automated.}  

\section*{Methods}
\mg{To study the opinion dynamics and the effect of the RDN, we employ two computational models of opinion dynamics: first, an activity-driven model  \cite{baumann2020modeling,baumann2021emergence} and a second, a selective social influence model \cite{sasahara2021social}. Both models lead to the formation of echo chambers. In the following, we describe the models in detail.  }
\subsection*{ \mg{Activity-driven model}}

We use an activity-driven model of opinion dynamics introduced in Baumann et al.\cite{baumann2020modeling,baumann2021emergence}. For a system of $N$ agents, each agent $i$ has an evolving opinion $x_i(t) \in [-\infty, \infty]$. For a given issue, agent $i$ has a stance with sign $\sigma(x_i)$ and a conviction with strength $|x_i|$. Strong convictions correspond to one of two extremes. Agent opinions change based on their activity-driven interactions with other agents $A_{ij}(t)$ \mg{(also referred to as the temporal adjacency matrix\cite{barabasi2005origin})}, the strength of social interactions $K > 0$, and the \mg{\textit{controversialness}} of the issue $\alpha > 0$ as in Refs \cite{baumann2020modeling,baumann2021emergence}. 

The opinion dynamics is given by 
\begin{equation}
\label{main.eq}
    \dot{x}_i= -x_i + K \left(\sum^{N}_{j=1} A_{ij} (t)  \tanh{(\alpha x_j)}\right) + D \mathcal{R}
\end{equation}
\mg{where $\mathcal{R}$ is the random dynamical nudge (RDN) term, with strength $D$}. If an agent with a set activity level $a_i \in [\varepsilon, 1]$ is active at time $t$, then they will interact with $m$ other agents, weighted by the probability $p_{ij}$ that agent $i$ would connect with agent $j$. These interactions are captured by the temporal adjacency matrix: $A_{ij}(t)=1$ when there is an input from agent $j$ to $i$ and $A_{ij}(t)=0$ otherwise. The probabilistic reciprocity factor $r \in [0, 1]$ determines the chance that a connection is mutually influential, $(A_{ij}(t)=A_{ji}(t)=1)$. If the interaction is reciprocated, both agents update their opinions; otherwise, only one of the agents' opinions is updated. 

The probability distribution of activities follows a power-law decay
\begin{equation}
\label{activities.eq}
  F(a) = \frac{1-\gamma}{1-\varepsilon^{1-\gamma}} a^{-\gamma}
\end{equation}
where $\gamma = 2.1$ governs the decreasing function's steepness of the activity probability distribution and $\varepsilon = 10^{-2}$ is the minimum activity. Most agents will have low activity and a few agents will have a high activity; most agents with low activity have little conviction; in contrast, active agents have greater conviction \cite{baumann2020modeling}. We define the connection probabilities as a function of the absolute value of opinion difference between two agents:

\begin{equation}
\label{conn.eq}
  p_{ij} = \frac{|x_i - x_j|^{-\beta}}{\sum_j{|x_i - x_j|^{-\beta}}}
\end{equation}

where $\beta$ is the homophily factor, the tendency for agents with similar opinions to interact with each other: $\beta = 0$ refers to no interaction preference, and $\beta > 0$ means agents with similar opinions are more likely to interact. Eq. \ref{conn.eq} is modeled as a power-law decay of connection probabilities with only a small chance for agents with opposite opinions to interact.

\mg{Note that an active agent $i$ interacts with is constant ($m =10$ unless otherwise stated), but whether an agent is active depends on their set activity probability ($a_i$, generated from Eq. \ref{activities.eq} at network instantiation). Furthermore, the connection probabilities, $p_{ij}$ (Eq. \ref{conn.eq}), is dynamic as opinions change over time. Together, the resulting interactions are captured by the temporal adjacency matrix $A_{ij}(t)$, which may be accumulated over time  - $\sum_t^T A_{ij}(t)$ - as an indication of which agents interacted with which other agents most frequently. The activity-driven model of opinion dynamics is thus defined by the network connectivity parameters $(\varepsilon, \gamma, m, \beta)$ and the issue parameters $(K, \alpha)$, with the RDN term $(D, \mathcal{R})$ being a novel addition in this work. We show the numerical procedure in Algorithm \ref{alg:cap}.}

\begin{algorithm}
\caption{Activity-driven model of opinion dynamics}\label{alg:cap}

\begin{algorithmic}
    \For{each agent $i$ in network of size $N$} \Comment{Initial conditions of the network}
        \State $a_i \gets \sim F(a)$  \Comment agent activities (Eq. \ref{activities.eq})
        \State $x_i \gets \sim U(-1,1)$  \Comment agent opinions
    \EndFor
    
    \While{$t < T $} \Comment{Simulation of opinions over time}
        \For{each agent $i$}
            \For{each agent $j$}
                \State $p_{ij} \gets \frac{|x_i - x_j|^{-\beta}}{\sum_j{|x_i - x_j|^{-\beta}}}$ \Comment{update connection probability (Eq. \ref{conn.eq})}
                \State $A_{ij}(t) \gets 0$ \Comment{temporal adjacency matrix}
            \EndFor
            \If{$a_i \geq \sim U(0,1)$} \Comment agent is active
               \State select $m$ other agents $j$ weighted by $p_{ij}$
               \For{each agent $j$ in $m$}
               \State $A_{ij}(t) \gets 1$ \Comment{agent $i$ interacts with agent $j$}
                \If{$r \leq \sim U(0,1)$}
                    \State $A_{ji}(t) \gets 1$ \Comment{reciprocal connection}
                \EndIf
                \EndFor
            \EndIf
        \State $\dot{x}_i \gets -x_i + K \sum^{N}_{j=1} A_{ij} (t) \tanh{(\alpha x_j)} + D \mathcal{R}$ \Comment update opinion (Eq. \ref{main.eq})
        \EndFor
        \State $t \gets t + dt$
        
    \EndWhile
\end{algorithmic}
\end{algorithm}

If active agents have an equal chance of interacting with $m$ other agents regardless of their stance ($\beta = 0$), then the network can become radicalized, with all agents having the same stance (Fig. \ref{fig1.adding}c and d).
If interactions are biased towards those with similar opinions, this can lead to the polarization of opinions and the formation of echo chambers, where most agents hold a moderate stance on a binary issue and few if any, agents have a neutral opinion\cite{baumann2020modeling} (Fig. \ref{fig1.adding}a and b). 

To quantify the effect of interventions on the distribution of opinions, we formulate a metric $\Lambda_x$ that gives the opinion-distance between the polarized peaks of the distribution. Formally, 

\begin{equation}
    \label{peak_dist.eq}
     \Lambda_x = \argmax_{x>0} \frac{f}{w} - \argmax_{x<0} \frac{f}{w}
\end{equation}

where $f$ is the frequency of opinions in a bin width of $w$, which was determined from the minimum of the Sturges \cite{sturges1926choice} and Freedman-Diaconis \cite{freedman1981histogram} bin estimation methods $w = \min ( \frac{\max x_R - \min x_R}{\log_{2}N + 1}, 2 \frac{\rm{IQR}}{N^\frac{1}{3}} )$ where $x_R$ is an opinion subset ($x>0$ or $x<0$) and $\rm{IQR}$ is the $x_R$ interquartile range.

The peak distance $\Lambda_x$ can be intuitively understood as the degree of polarization of echo chambers. For $\Lambda_x$ close to 0, the distribution of opinions is normal and depolarized, and for larger $\Lambda_x$, the opinions of agents are polarized (Fig. \ref{fig1.adding}f). %

\subsection*{\mg{Selective social influence model}}
\mg{To ensure that the results are not specific to a given model of opinion dynamics, we studied the effects of the RDN on a selective social influence model of opinion dynamics that leads to the formation of echo chambers \cite{sasahara2021social}.
}
\mg{
In brief, the selective social influence model evolves by selecting an agent $i$ at time $t$ (only one agent is chosen per time step) and showing a screen of $m$ recent messages from other agents $j$. An agent's opinion $x_i$ changes according to}
\begin{equation}
    \label{social_influence.eq}
    \dot{x}_i= -x_i + \mu \frac{\sum_{j=1}^{m} I_{\epsilon} (x_i,x_j) (x_j-x_i)}{\sum_{j=1}^{m} I_{\epsilon} (x_i,x_j)}
\end{equation}
\mg{
where $\mu$ is the influence strength parameter and $I_{\epsilon}$ is an indicator function for concordant opinions bounded by $\epsilon$:
}
\begin{equation}
    \label{concordant.eq}
    I_{\epsilon} (x_i,x_j) = 
    \begin{cases}
        1, & \text{if } |x_i - x_j| < \epsilon \\
        0, & \text{otherwise}
    \end{cases}
\end{equation}

\mg{
In addition, the selected agent posts a message: either a re-post of a concordant message ($|x_i - x_j| < \epsilon$) or a new message with the agent's new opinion (with probabilities $p$ and $(1-p)$, respectively).
}

\mg{
Lastly, the agent at time $t$ has a probability $q$ of rewiring its connections. Here, agents can replace the connection with a randomly chosen ones. Other strategies of selecting a new friend were explored in the Ref \cite{sasahara2021social} and exacerbated echo chamber formation\cite{sasahara2021social}. The number of peaks was calculated as in Ref\cite{peaks_2017}. We extended the selective social influence model of echo chamber formation by adding $D\mathcal{R}$ (according to Eq. \ref{RDNunique.eq}) to Eq. \ref{social_influence.eq}. 
}

\section*{Results}

\begin{figure}[ht!]
\centering
    \includegraphics[width=1\linewidth]{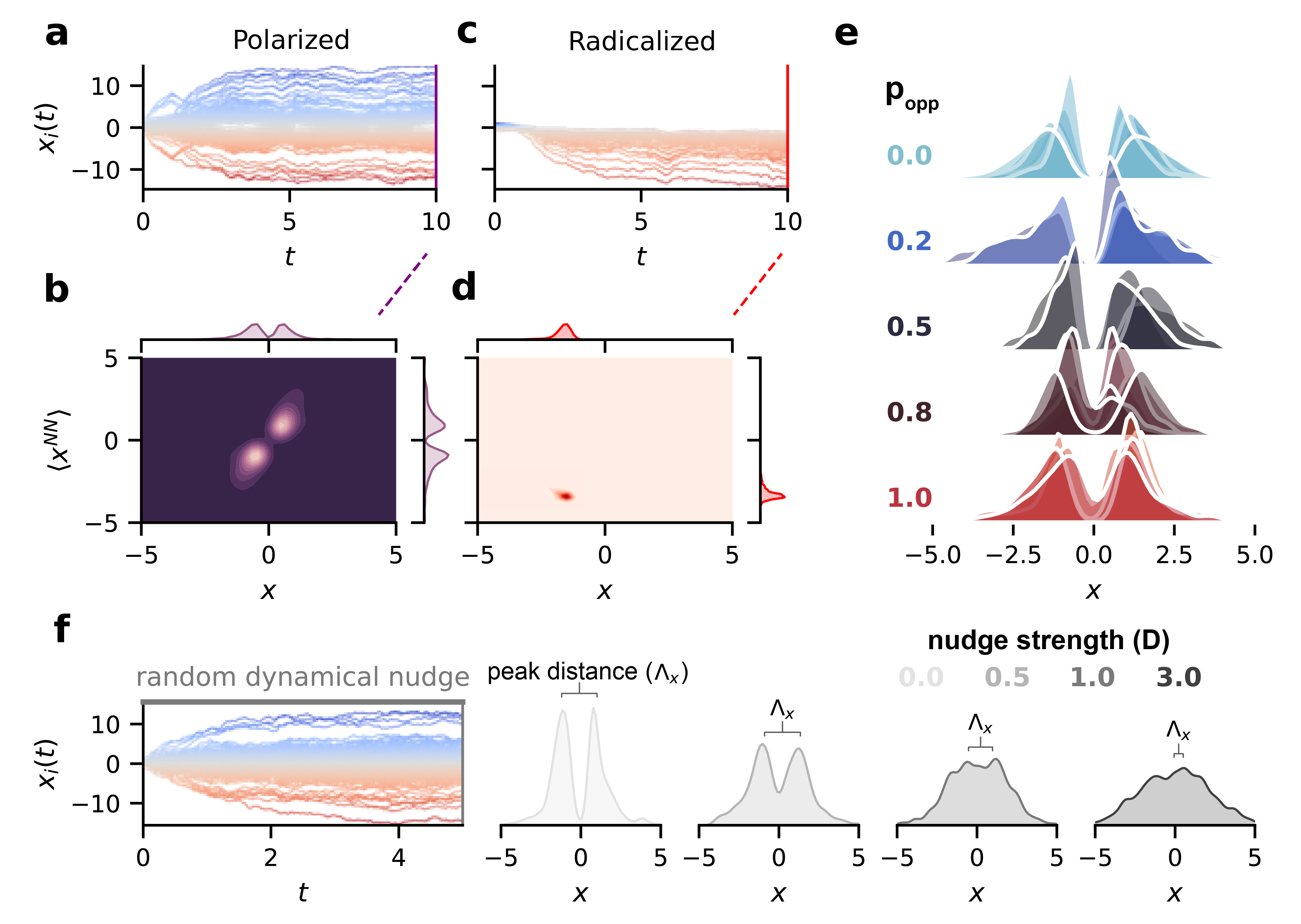}
    \caption{
        \textbf{Adding a random dynamical nudge prevents a network from becoming polarized.} 
        The time traces of $N=1000$ agents for (a) polarized ($\beta=3$) and (c) radicalized ($\beta=0$) states. (b, d) At $t = 10$, heat maps of agents' opinions versus mean of agents' nearest neighbours ($\langle x^{NN} \rangle$) to indicate the formation of echo chambers - agents interact with other agents who have similar opinions - and the resulting bimodal distribution of opinions for polarized networks. (e) $p_{opp}$ represents the probability of randomly flipping the homophily factor ($\beta$) to instead interact more with those holding opposite opinions. $p_{opp}$ did not appear to be effective and instead many simulations produced radicalized networks by $t=10$ (5 simulations individually plotted per $p_{opp}$). We introduce a "random dynamical nudge" (RDN, f) that prevents a network of agents from becoming polarized with sufficient strength $D$. The peak distance $\Lambda_x$ is shown by the width of the brace for each $D$. Here, the RDN is the Wiener process $\xi(t)$. The degree of polarization is indicated by the peak distance ($\Lambda_x$). Other simulation parameters were $m=10$, $K=3$, $\alpha=3$, and $r=0.5$.
    }
    \label{fig1.adding}
\end{figure}

Intuitively, randomly negating the homophily factor $\beta$ for some agents should prevent the bimodal distribution by increasing the chance an agent would interact with others holding opposing opinions. However, the increased tendency for some agents to interact temporarily more with those of opposing opinions is not enough to change the overarching network dynamics. Regardless of the probability of interacting with the opposite opinion, $p_{opp}$, we found that the echo chambers remain in place (Fig. \ref{fig1.adding}e). In many cases, the networks became radicalized or even more polarized instead. This may be due to the diminishing influence of $\beta$ on interactions as $p_{opp}$ increases such that a network with $p_{opp} = 0.5$ is similar to that with $\beta = 0$. That is, agents no longer interact with others of similar opinions but effectively randomly interact with other agents as in $\beta = 0$, which has been established to lead to a radicalized state.

Simply increasing the probability of interaction with agents with opposite opinion was not sufficient to depolarize the echo chambers. Next, we sought to impose an additional term that may accomplish this. Our approach was to include a random dynamical nudge (RDN, $\mathcal{R}$) term with strength $D$ to prevent a network of agents from becoming polarized \mg{(see Eq. \ref{main.eq})}. In the limit of $D \rightarrow 0$ the effect of the nudge diminishes.

Our first approach was to have the nudge term be the Wiener Process $\xi(t)$ \mg{due to its nature of continuously generating independent, zero-centered, normally-distributed values \cite{RICCIARDI1976185, zbMATH00192908}}(Fig. \ref{fig1.adding}f). \mg{Formally, between 2 points, $\xi(t)$ and $\xi(t + s)$, the distribution of values is $\sim \mathcal{N}(0, s)$}. Although this accomplished the goal of depolarizing the network, it is not a feedback mechanism. \mg{Due to the independent nature of $\xi(t)$ to the opinions in the network, we consider this an external mechanism. For social networks, a mechanism of noise that uses properties of the network itself (i.e. internal to the network), therefore, has a more plausible implementation than the external noise generated by $\xi(t)$. }



\begin{figure}[ht!]
\centering
    \includegraphics[width=0.95\linewidth]{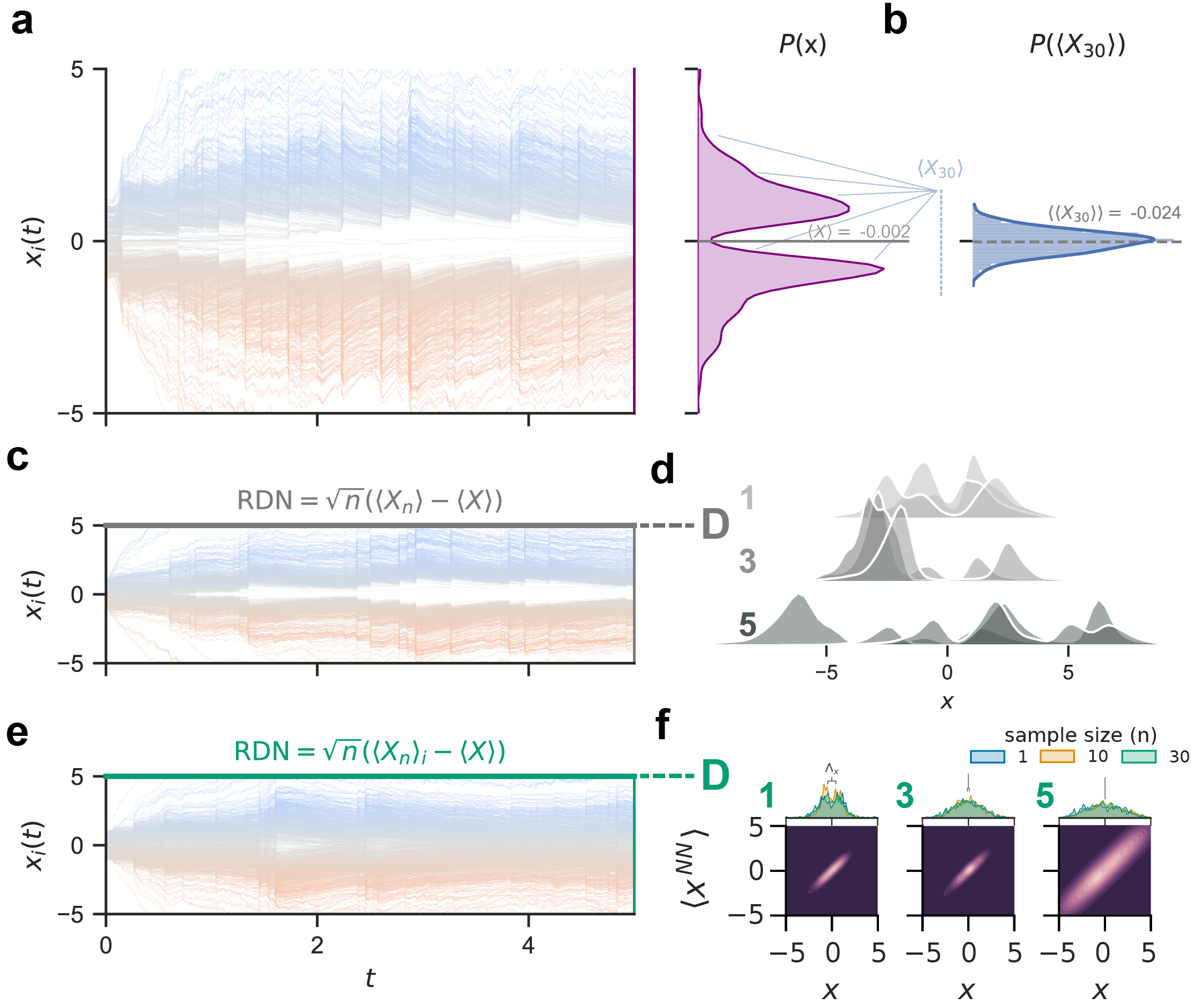}
    \caption{
        \textbf{The normalizing effect of a random dynamical nudge based on the central limit theorem. } 
        Given a polarized network of agents (a), the mean of the opinions $\langle X \rangle$ at $t=5$ is close to 0. An approximation of this mean value, using the Central Limit Theorem (CLT), was gathered by sampling 30 opinions $\langle X_{30} \rangle$ (i.e. with sample size $n$ of 30) a thousand times at $t=5$ and taking the mean of these sample means: $\langle \langle X_{30} \rangle \rangle$ (b). Applying a RDN formulated as a CLT term, but where each agent sees the same sample mean $\langle X_n \rangle$, was insufficient to achieve a normal distribution of agent opinions (c), regardless of RDN strength $D$ (d). Instead, each agent must see their own sample mean  $\langle X_n \rangle_{i}$ to prevent a network from becoming polarized (e). This was true for different RDN strengths ($D \in \{0, 1, 3, 5\}$). Varying the sample size $n$ did not significantly change distributions. The RDN causes the opinions to be predominantly neutral ($\approx 0$) and prevented the formation of echo chambers. Large RDN values (e.g. $D=5$) also have more agents with extreme opinions. Other parameters were $N=1000$, $K = 3$, $\beta=3$, $\alpha=3$, $m=10$, $r=0.5$, $T=5$, and $n=30$ unless otherwise stated.
    }
    \label{fig2.hists}
\end{figure}

To build an effective RDN term, motivated by the Lindeberg–Lévy central limit theorem (CLT)\cite{Billingsley1961TheLT}, we sampled opinions from the network in such a way to obtain a normal distribution of opinions (Fig. \ref{fig2.hists}a and b). In addition, the mean of the Lindeberg–Lévy CLT is 0, which is a desirable property for \mg{reaching a neutral consensus}. As our starting point for a socially plausible nudge we formulate the RDN as $\sqrt {n}\left( \langle X_{n} \rangle -\langle X \rangle \right)$ (Fig. \ref{fig2.hists}c),
where $\langle X_{n} \rangle$ is the mean from a sample (of size $n << N$) of opinions, and $\langle X \rangle$ is the true mean of all opinions.

In applying this RDN, we find the mean of a random sample of opinions (that is, $\langle X_{n} \rangle$) at each time step $dt$ and show every agent the \textit{same} $\langle X_{n} \rangle$ (Fig. \ref{fig2.hists}c). However, this can lead to radical or asymmetrical opinion dynamics as the population's opinions were pulled towards a single sample mean at each $dt$ (Fig. \ref{fig2.hists}d). This is similar to previous work\cite{perra2019modelling} with a central influencer pulling the network dynamics towards a point. To depolarize a population, we instead hypothesized that \textit{each agent} would need a \textit{unique} sample of opinions ($\langle X_{n} \rangle_i$). That is, where

\begin{equation}
\label{RDNunique.eq}
    \mathcal{R} = \sqrt {n}\left( {\langle X_{n} \rangle}_i -\langle X \rangle \right)
\end{equation}

The addition of this RDN term to the opinion dynamics (Eq. \ref{RDNunique.eq}, Fig. \ref{fig2.hists}e) shifted the collective opinion from an expected bimodal distribution to a unimodal distribution (Fig. \ref{fig2.hists}f). Even with RDN strength $D$ of 1, there were many more neutral opinions than without RDN. With $D=3$, the opinions were depolarized with a median of $\approx$ 0. However, stronger RDNs created an undesirable paradoxical effect of increasing the number of opinions with strong convictions, i.e., increasing the standard deviation of the opinions and spreading the distribution. Interestingly, the effect of sample size ($n$) was negligible. The lack of influence from the sample size was likely due to the uniqueness of samples being sampled at every $dt$ approximating a similarly normal distribution. The consequence of this is that a reduced RDN formulation may be possible and was indeed explored later in Table \ref{RDN.table} and Fig. \ref{robustness.fig}. For now, we consider the foundational case of the RDN as formulated in Eq \ref{RDNunique.eq} with $n=30$ for the rest of the results.

We investigate the effect of the RDN on echo chamber formation by viewing agents' nearest neighbors in the network (Fig. \ref{fig2.hists}f). The two echo chambers typically formed at $D=0$ (see Fig. \ref{fig1.adding}b) were instead neutralized into a single community with $D \geq 1$. Although neighbors for $D>=1$ still tended to have a similar opinion, there were more agents with a neutral opinion that facilitated further communication across stances on an issue. Again, large $D$ could spread the community's opinion distribution, which means more agents have strong convictions even as many agents remain neutral. Hence, the RDN strength $D$ plays an essential role in the distribution of opinions.

Fundamentally, the RDN emulates a social network showing each person a \textbf{random} independent sample of peers' opinions on a topic. If implemented appropriately, it represents a tractable approach to prevent echo chambers in real social media.

\begin{figure}[ht!]\centering
    \includegraphics[width=0.7\linewidth]{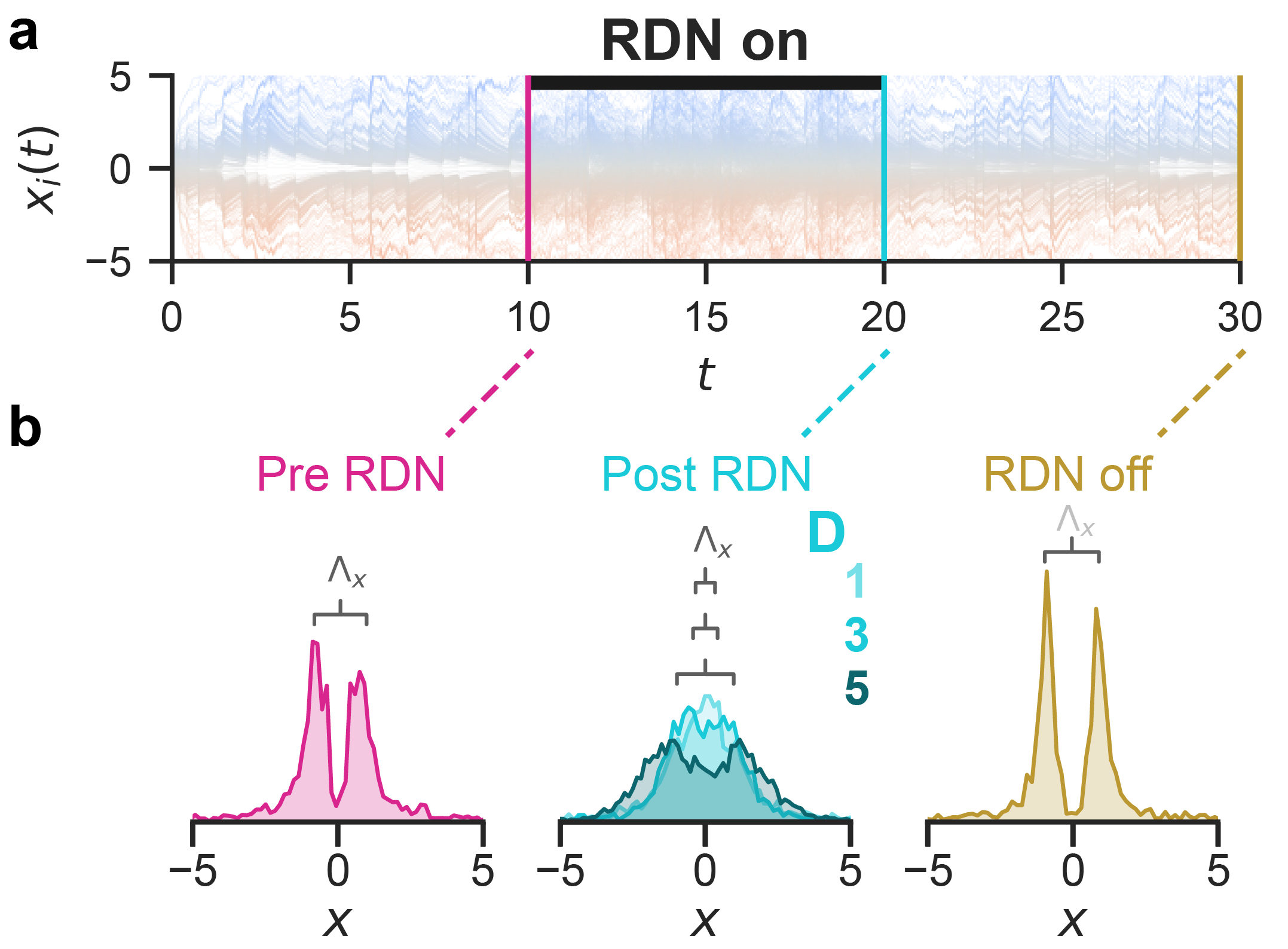}
    \caption{
        \textbf{The depolarization of existing echo chambers by a RDN.} 
        The opinions of 1000 agents over time (a) without the RDN until $t=10$, with the RDN until $t=20$ ($D=3$, black bar), and finally without the RDN until $t=30$.
        Kernel density estimates of agent opinions (5 trials each) at moments in time (b) at $t=10$ ("Pre RDN", pink), $t=20$ ("Post RDN", $D \in {1,3,5}$, blue), and $t=30$ ("RDN off", tan). The peak distance ($\Lambda_x$) decreases when a RDN is applied.
        Other parameters were $m=10$, $K=3$, $\alpha=3$, $\beta=3$, $n=30$, and $r=0.5$. 
    }
    \label{fig3. depolarize}
\end{figure}

The results thus far have shown that a RDN can prevent the polarization of an opinion dynamics system. Next, we assess the efficacy of a RDN in depolarizing a network that has already formed echo chambers (Fig. \ref{fig3. depolarize}). We allowed a system to become polarized until $t=10$ and then applied a RDN for an equal time ($t=20$; Fig. \ref{fig3. depolarize}a). We also examined the after-effects of applying a RDN by removing it again ($t=30$). By examining the opinion distribution at different time points (Fig. \ref{fig3. depolarize}b), we found that adding the RDN depolarized the system. Removing the RDN can cause the system to regress quickly, so the RDN should remain as part of the system. 
This result shows that our proposed RDN not only prevents the formation of echo chambers but also depolarizes existing ones.

\begin{figure}[ht!]\centering
    \includegraphics[width=1\linewidth]{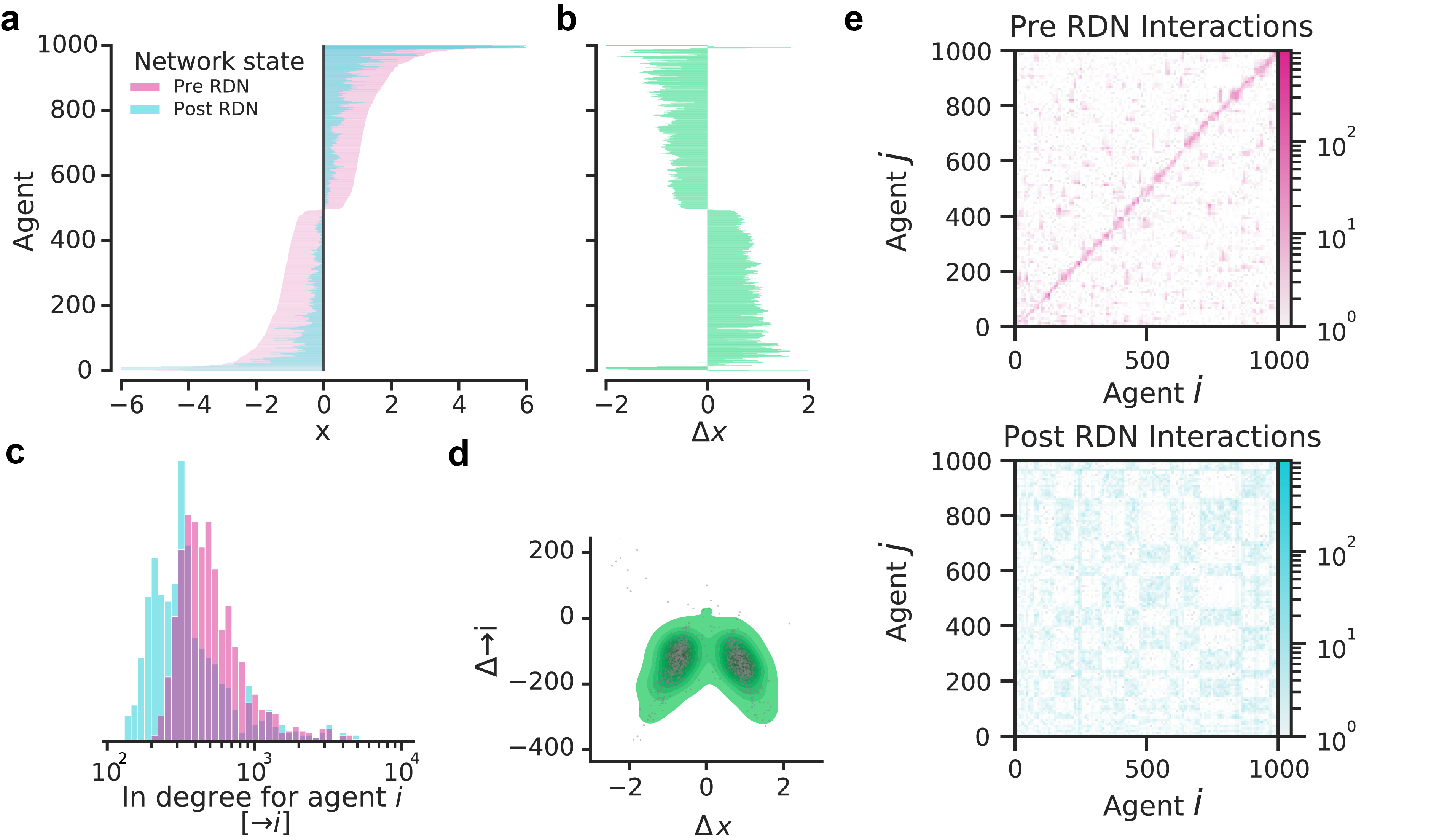}
    \caption{
        \textbf{The effect of RDN on an agent's opinion and interactions in the network.} 
        From a simulation in Fig. \ref{fig3. depolarize} ($D=3$), each agent $i$'s opinion $x$ is shown at $t=10$ ("Pre RDN", pink) and at $t=20$ ("Post RDN", light blue) (a), as well as their change in opinion, $\Delta x$, (b). 
        The incoming interactions from other agents to an agent $i$ ("in degree", $\rightarrow i$) is shown in (c) as a probability histogram $P(\rightarrow i)$.
        The density of in degree change $\Delta \rightarrow i$ as a function of $\Delta x$ (d).
        The interactions between agents before the RDN (e, top panel) show a large number of interactions and small groups. That is in contrast to after the RDN (e, bottom panel), which shows fewer interactions but larger groups. Note that activity probabilities $a_i$ remained the same throughout the simulation. Other parameters were $m=10$, $K=3$, $\alpha=3$, $\beta=3$, $n=30$, and $r=0.5$.
    }
    \label{fig4.network}
\end{figure}

For further network analysis, we looked at a network like that in Fig. \ref{fig3. depolarize} and explored the depolarizing effect of the RDN on each agent $i$ based on their opinion $x_i$ (Fig. \ref{fig4.network}). The various degrees of opinion change $\Delta x$ indicated that the effect of the RDN was not homogeneous across users (see Fig. \ref{fig4.network}a and b). There are diminishing returns to the RDN, which is to be expected. Users closer to $x=0$ are already less polarized and, thus, are more likely to engage with other agents. On the other hand, users far away from $x=0$ depolarize more. These agents are less likely to be engaging with opinions dissimilar to their own, and being exposed to the RDN should have a greater effect on moving them towards the center. Note, however, that the most polarized users are harder to sway and the RDN is less effective. At the very extremes of the distribution, the RDN has an opposite effect.

Finally, we look at the topography of the network to get a glimpse at the mechanisms driving the effects of the RDN. Similar to social media networks, the nodes in our simulated networks are individuals, and the edges between nodes are created when there is an interaction among individuals. In a Twitter network, for example, each user is a node and an edge is created when a user $H$ retweets user $A$. In this layout, the author of the original tweet is the authority ($A$) and the author of the retweet is set as the hub ($H$), such that $H_{retw} \rightarrow A_{tw}$. Social media networks, as well as our simulated network, have highly skewed in-degree distributions, where few, highly visible, users produce most of the information that is reproduced \cite{calvo2020fake,aparicio2015model}.

Figure \ref{fig4.network}c shows that activating the RDN changes the distribution of in-degrees in the network. The change suggests that, once agents are exposed to other opinions, individuals will engage less intensely with other users. We also found that the amount of in-degree change is reflected in the change of opinion (see Fig. \ref{fig4.network}d). However, while the number of intense interactions decreased, there was an increase in the diversity of interactions (see Fig. \ref{fig4.network}e). The RDN is moving the opinion of users by increasing the variety of interactions while also decreasing the number of interactions with individual users (see Fig. \ref{fig4.network}e). This points to a less intense, yet more plural, engagement.

The current RDN formulation in Eq.\ref{RDNunique.eq} relies on the true mean $\langle X \rangle$, something that would be hard to measure in practice. To investigate a more practical implementation of RDN by stripping it to its bare essentials, we ran extensive simulations with different RDN terms (see Table \ref{RDN.table} and Fig. \ref{robustness.fig}a). 

\begin{table*}[ht]
\caption{Different RDN terms yielding similar results}
\begin{adjustbox}{max width=17.5cm}
\begin{tabular}{|c|c|lll}
\cline{1-2}
\textbf{RDN} & \textbf{Meaning.} \small{Note that each agent $i$ is shown their own RDN.} \\
  \cline{1-2}
$D \sqrt {n}\left(\langle X_{n} \rangle_i - \langle X \rangle \right)$ & sample mean of opinions compared against true population mean of opinions and scaled by sample size  \\
  \cline{1-2}
$D \left(\langle X_{n} \rangle_i - \langle X \rangle \right)$& sample mean of opinions compared against true population mean of opinions  \\
  \cline{1-2}
$D \left( X_{1}  - \langle X_{n} \rangle \right)_i$& sample agent's opinion compared against sample population mean of opinions  \\
  \cline{1-2}
 $D \left( X_{1} - X_{1}' \right)_i$& sample agent's opinion compared against another sample agent's opinion  \\
  \cline{1-2}
 $\pm D \cdot X_{1_i} $ & a single sample agent's opinion  \\
 \cline{1-2}
\end{tabular}
\label{RDN.table}
\end{adjustbox}
\end{table*}

In all setups, adding the RDN was beneficial in depolarizing the echo chambers and, given the optimum value of $D$, RDN leads to a uni-peak distribution of opinions with a mean around the neutral consensus. Fig.\ref{robustness.fig}a shows the distribution of the collective opinions for different RDN terms introduced in table \ref{RDN.table}. We note that even showing a single random opinion to each agent is highly beneficial in reducing the polarization. The critical part of the RDN was that each agent $i$ was shown another random agent's opinion with a weight at least as strong as the usual interaction specified in the original equation. 
That is, the effects of a homophilic system ($\beta>0$) were overcome by focusing each agent on a random opinion from the system. This was the case even in the extreme scenarios with many social interactions $K=3$ for a controversial topic ($\alpha=3$) as shown in Fig. \ref{robustness.fig}a. Remarkably, no extra restrictions were required, such as requiring reciprocity or being of the opposite stance to depolarize the agents' opinions.

Next, we sought to determine how sensitive the dynamics were to the RDN strength $D$ (Fig. \ref{robustness.fig}b). Increasing RDN strength $D$ tends to decrease $\Lambda_x$ up to an optimal strength that depends on $\alpha$ and $K$. More controversial systems and those with stronger social interactions require a stronger RDN; $D$ therefore needs to be considered on a case-by-case basis. Note that homophilic networks $\beta>0$ have similar $\Lambda_x$. This is similar to the coherence resonance observed in dynamical systems where an optimum amount of noise intensity leads to minimum variability \cite{neiman2007coherence}.

\begin{figure}[t!]\centering
    \includegraphics[width=0.9\linewidth]{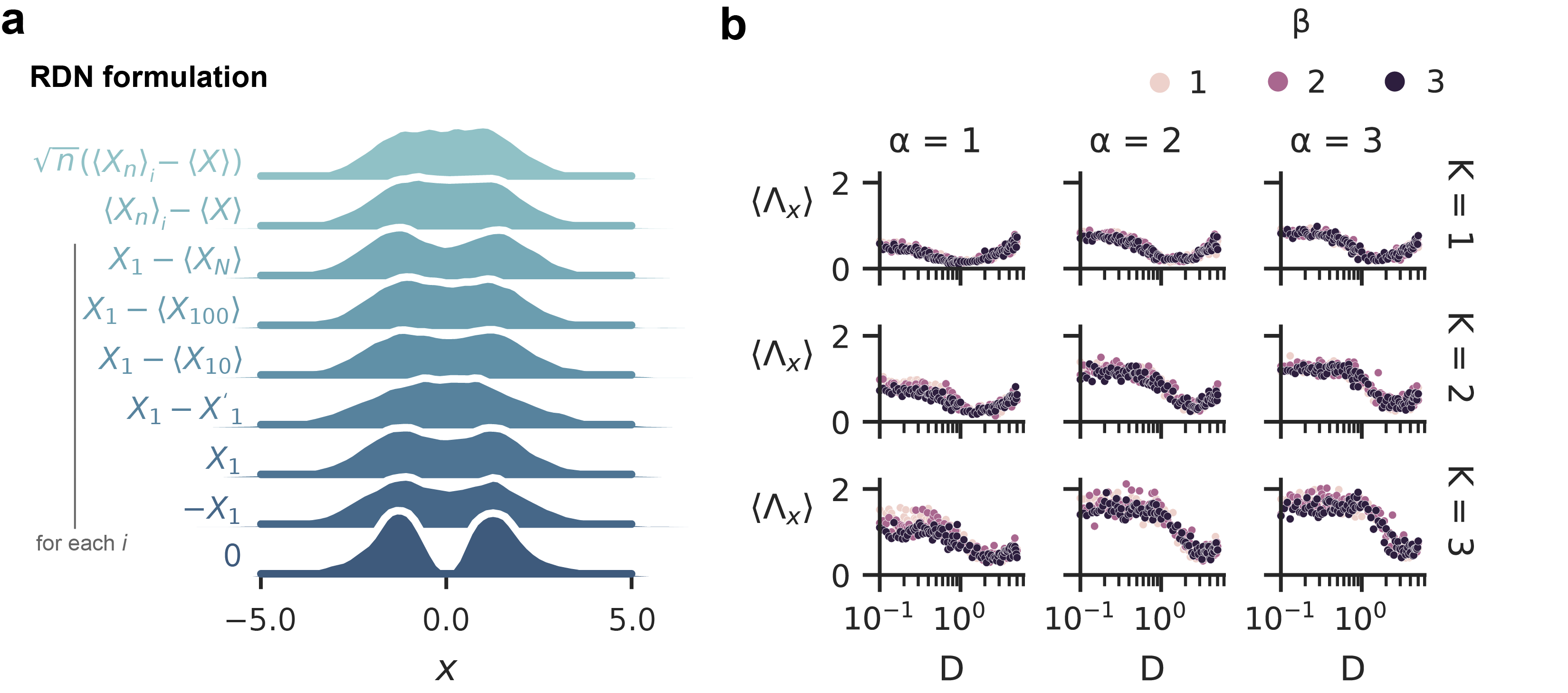}
    \caption{
        \textbf{Assessing the robustness of the RDN}. 
        (a) Different RDN formulations produce similar results as long as each agent gets a different sample ($D=3$, 20 trials). $D=0$ is shown in the last row for comparison. See \ref{RDN.table} for equation explanations.
        (b) The scatter plots in each panel shows the polarized peak distance $\Lambda_x$ averaged over sample size $1... 50$. Columns indicate different values of $\alpha$. Rows indicate different values of $K$. Colors indicate different values of $\beta$. 
        Other parameters were $N=1000$, $m=10$, $r=0.5$, and $T=5$.    
    }
    \label{robustness.fig}
\end{figure}

\mg{Finally, we consider the RDN applied to another recent model of echo chamber formation, namely that of selective social influence \cite{sasahara2021social}. Specifically, the model incorporates sharing of $m$ messages between $N$ agents in a partially connected network with $E$ edges. Messages are shared in the network with probability $p$ of being reposted if they are concordant with an agent's opinion (according to $|x_i - x_j| < \epsilon$, Eq. \ref{concordant.eq}), or an agent instead shares their own opinion in a new message with probability $1-p$. An agent's opinion and connections change based on message agreement: concordant messages selectively influence an agent's opinion (up to $\mu$, Eq. \ref{social_influence.eq}), and discordant messages selectively change an agent's connections (with rewiring probability $q$). 
}

\begin{figure}[t!]
\centering
    \includegraphics[width=1\linewidth]{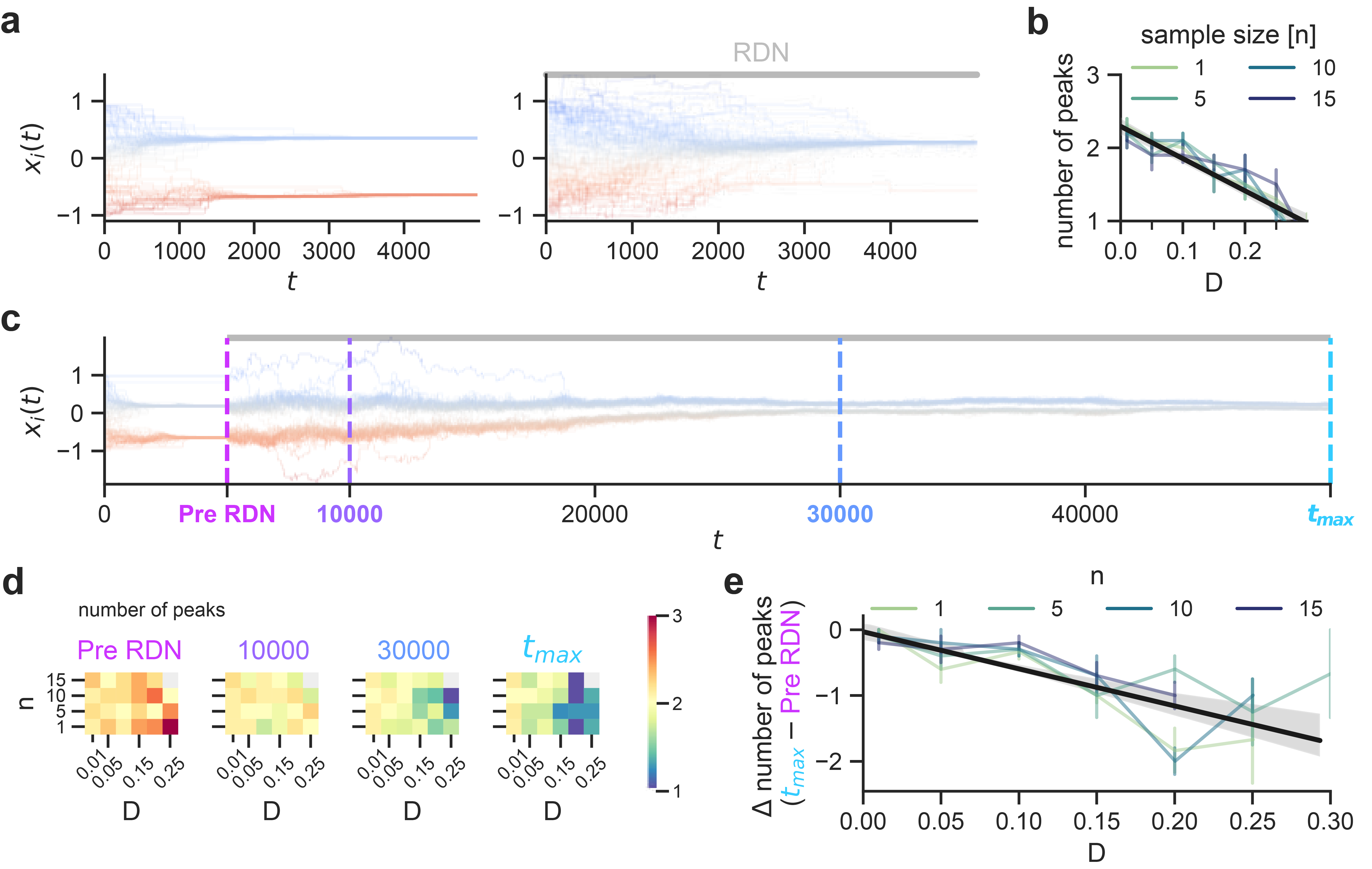}
    \caption{
        \textbf{The RDN applied to a selective social influence model of echo chamber formation.} 
        \mg{A different opinion dynamics model that typically produces echo chambers (a, left)\cite{sasahara2021social} can prevent these echo chambers from forming by applying the RDN as in Eq. \ref{RDNunique.eq} (a, right). Both simulations in (a) have the same seed conditions.
        (b) The number of peaks formed after a 5000 time step simulation was dependent on the RDN strength $D$, but not clearly on sample size $n$ (10 trials each, bars indicate standard error of the mean). Note that the selective social influence model can produce more than 2 echo chambers.
        (c) Applying the RDN ($D=10$ and $n=1$) to an already-polarized network "Pre RDN" took longer to become depolarized, if at all ($t_{max} = 50000$). 
        (d) The number of peaks over time for a range of $n$ and $D$ (average of 10 trials). By $t_{max}$, most networks had stabilized, with larger D values causing fewer peaks. This change in the number of peaks between "$t_{max}$" and "Pre RDN" is shown in (e). Note that for $D>=0.3$ the simulations could become numerically unstable (any change in opinion of $>1$ within 10 time steps) and were excluded (grey block in "d" indicates less than 3 stable simulations).
        RDN in examples (a, c) have strength ($D$) of 0.2 and sample size ($n$) of 1.
        Other parameters were as in the authors' original model: $N=100$, $E=400$, $m=10$, $\epsilon = 0.5$, $\mu =0.5$, $p=0.5$, and $q=0.5$.
        }
    }
    \label{fig6.social_influence_model}
\end{figure}

\mg{We found that applying the RDN could be an effective approach for preventing echo chamber formation and could depolarize existing echo chambers over time. (Fig. \ref{fig6.social_influence_model}). Instantiating a specific network instance that would normally produce 2 opinion peaks or echo chambers ($N=100$, $E=400$ $m=10$, $\epsilon = 0.5$, $\mu =0.5$, $p=0.5$, $q=0.5$, Fig. \ref{fig6.social_influence_model}a), has a single peak with the RDN ($D=0.2$ and $n=1$). In agreement with previous results, stronger RDNs caused fewer peaks (Fig. \ref{fig6.social_influence_model}b), but sample size seemed to be negligible.
}

\mg{Applying the RDN ($D=0.2$ and $n=1$) to an already polarized network ($t = 5000$, "Pre RDN", Fig. \ref{fig6.social_influence_model}c and d) was also effective, but took much longer to reduce the number of opinion peaks ($t_{max} = 50000$). Comparing the number of peaks at $t_{max}$ versus at "Pre RDN" time points again revealed the dependence of the resulting networking on the RDN strength.
}

\mg{Although larger RDN strengths were generally better, the simulations with a delayed RDN (like Fig. \ref{fig6.social_influence_model}c) could become numerically unstable - have an increasingly large range of opinions - when $D>=0.3$. This was likely due to the relationship between the influence strength $\mu$ and RDN strength $D$; where $D>\frac{\mu}{2}$ may be problematic due to $D$'s large influence. However, because this primarily seems to arise when applying the RDN to a polarized network, the bounded confidence parameter $\epsilon$ may also be a factor. Further investigation is warranted for a fine-grained understanding of this phenomenon, including the use of alternative numerical solvers.
}

\section*{Discussion}

Many studies report the formation of echo chambers in social media \cite{baumann2020modeling,baumann2021emergence,cinelli2021echo,jamieson2008echo,sasahara2019inevitability}, whereby agent's interactions lead to the formation of two segregated groups with minimal or no interaction between them. On issues with an expertly-informed side, such as climate change, vaccines, LGBTQ+ rights, etc., only one side is supported by facts. In topics where there is no correct position, such as during most elections, polarized opinions can lead to incivility\cite{gervais2019rousing} and animosity\cite{webster2017ideological,rogowski2016ideology}. For such events, more exchange and greater communication across groups lead to more democratic deliberation\cite{morrell2018listening} and consensus-building\cite{olmastroni2020deliberative,susskind1999consensus,niemeyer2007ends}. The question that we aim to answer is what mechanism can prevent the formation of echo chambers in such events?

Activity-driven models can capture the formation of echo chambers in social media, accurately reproducing the divide in opinions emerging from controversial issues\cite{perra2012activity,baumann2020modeling}. We use a modified activity-driven model where each agent receives a random input based on the opinions of the others in the network and counteracts the effects of homophilic interactions. We show that the random nudge stops the formation of echo chambers or disbands echo chambers already formed, and leads to a distribution of opinions centered around a shared central space. \mg{Of particular importance for a possible application of our RDN is that it depolarizes echo chambers by employing a user-agnostic nudge term. In other words, the RDN is independent of each agent's opinion, and it adds a new perspective by providing input from opinions out of an agent's immediate circle of interactions. Thus, our RDN does not require surveillance of every person's opinions.}

Furthermore, we also find heterogeneous effects of the RDN across the network, indicating that, in addition to depolarizing the network as a whole, the RDN depolarizes individual users across different network topographies at different rates. This also points to potential drawbacks to our intervention. Primarily that, at the very extremes of the distribution, the RDN might have the opposite effect on users, further polarizing their views. This last effect is in line with recent experimental research,\cite{bail2018exposure} where exposing some viewers to opposing ideologies might increase political polarization. There, unlike our RDN, researchers used messages from high-profile political elites (instead of randomly chosen opinions)\cite{bail2018exposure}. 

The model we built upon is relatively simple yet has been tested against empirical Twitter's data \cite{baumann2020modeling,baumann2021emergence}, and its main feature is that echo chambers transiently occur through interactions among the agents. Taken together, these results show how an activity-driven network model of opinion dynamics, which normally becomes polarized, can be depolarized with the addition of noisy feedback: the RDN. Our results suggest that the RDN a) prevents a network from forming echo chambers, b) can depolarize a network that already has echo chambers, c) that this facilitates interactions with more users, and d) that the effect is quite robust but depends on the issue at hand.

In Fig. \ref{fig1.adding} we show that our model leads to the formation of polarized groups where the transient echo chambers emerge due to interactions among the agents. Consistent with previous studies \cite{bail2018exposure}, simply showing the opposite opinion does not lead to the depolarization of echo chambers (Fig. \ref{fig1.adding}e). Bail et al.\cite{bail2018exposure} conducted a large field experiment where U.S. Twitter users were exposed to messages from opposing political ideologies. The results from the experiment show that exposing users to opposing views actually increases political polarization. However, in that study, the messages used came from high-profile political elites. Our RDN model exposes users to messages from other, randomly selected opinions, which might explain why, for most user, our intervention has the desired, depolarizing effect. Thus, the RDN might depolarize the network precisely because it comes from a myriad of sources, not a static partisan opposite.

To formulate a RDN that is plausible for a social network to implement, we randomly sampled opinions from the network and estimated the average (see Fig. \ref{fig2.hists}a and b). As an initial attempt, each agent was shown the same sample mean, but this still led to polarized or radicalized networks (Fig. \ref{fig2.hists}c and d). We found that to prevent a network from becoming polarized, agents must each be shown a \textit{unique} sample mean of random opinions. Of course, the strength of the nudge matters, and a nudge with vanishing strength does not affect the overall dynamics of the network.

Even when the network is highly polarized, including the nudge, $\mathcal{R}$, leads to the depolarization of the opinion dynamics network (Fig. \ref{fig3. depolarize}). After removing the nudge, it still has a brief after-effect and the system does not revert immediately to a polarized state. Fig. \ref{fig3. depolarize} shows the distribution of the opinions at three different points, before, during, and after the intervention. The lasting, if brief, effect of the RDN after removal can be desirable since it suggests that the nudge could be applied intermittently.

We show that including the nudge effectively changed interaction among agents leading to more diverse and less intense interactions as shown in Fig. \ref{fig4.network}. Agents change their opinions to become more, but not entirely neutral. Interestingly, and again in agreement with experimental observation\cite{bail2018exposure}, those with extreme opinions remain polarized and even move further toward their extreme when exposed to the nudge (Fig. \ref{fig4.network} a and b). 
\mg{It is important to mention that the RDN can encourage more diverse connections but that having random connections is not sufficient for a network to be depolarized. Specifically, equally ($\beta = 0$, Fig. \ref{fig1.adding}b) or randomly (e.g. $p_{ij} \sim U(0,1)$, not shown) generated probability connections produce radicalized opinion distributions. The intuitive reasoning is that equal and random interactions without the RDN leads to a winner-takes-all approach of the most active agents pulling the opinions of the network toward one stance\cite{baumann2020modeling}. The RDN term, however, causes opinions to be distributed around the middle-ground by design. Encouraging random connections may then not be sufficient to depolarize echo chambers. Instead, showing how an opinion compares to the middle-ground, or providing aggregated opinions, should be emphasized over the artificial adjustment of connections and may lead to more diverse connections. In brief, random connections do not appear to lead to diverse opinions, but the random aggregation of opinions seems to encourage diverse connections.}

Although the RDN in Eq. \ref{RDNunique.eq} was highly effective in inducing a consensus state, we did evaluate alternate versions (Table \ref{RDN.table}) that produced qualitatively similar results (Fig. \ref{robustness.fig}a). The effectiveness of even single random opinions, but strongly weighted, at nudging the network towards consensus indicates that the uniqueness of the nudge is a major influence in its efficacy. By nudging every agent's opinion dramatically and repeatably, this breaks down the homophily of the network. However, a nudge without a comparison may be more susceptible to becoming radicalized. For asymmetric conditions where the mean opinion is not zero, using the RDN (Eq. \ref{RDNunique.eq}) provides a robust approach. Furthermore, we evaluated the robustness of the nudge to $\alpha$, $K$, $\beta$, and $D$ (Fig. \ref{robustness.fig}b). We found that the peak distance $\Lambda_x$ was dependent on the controversialness of the issue ($\alpha$) and interaction strength of the agents ($K$), but not the degree of homophily (for $\beta  > 0$). The RDN strength ($D$) thus had differential effects that depend on the network structure. In general, stronger $D$ reduces $\Lambda_x$, but there is an optimum value of $D$ for a given network. For $D$ above this value, the opinion distributions become more diffuse with a greater spread of the data. This could be seen as weighting the noise's standard deviation too much, or more concretely, giving too much credence to all presented nudges. Consequently, real implementation of the RDN should be carefully curated and introduced gradually.

\mg{The depolarizing effect of the RDN is general enough to apply to both an activity-driven model\cite{baumann2020modeling} (Eq. \ref{main.eq}) and a selective social influence model\cite{sasahara2021social} (Eq. \ref{social_influence.eq}), both in which the authors performed empirical validation of their respective models. 
The selective social influence model is able to have a single peak in the original implementation if the agents' bounded confidence distance $\epsilon$ was high enough (>= 0.6). However, this is like a "tolerance" of others' opinions and thus is hard to change in an environment like Twitter. Furthermore, we chose the "random" selection of new followers as this was the slowest approach for forming echo chambers compared to following a recommended concordant agent ("recommendation") or following a re-posted message's originator ("re-post"). We chose the default parameters in Ref \cite{sasahara2021social} ($N=100$, $E=400$ $J=10$, $\epsilon = 0.5$, $\mu =0.5$, $p=0.5$, and $q=0.5$), which consistently produced at least two peaks and sometimes three.
}

\mg{
A few notable differences exist between the activity-driven model and selective social influence model. First, the activity-driven model gives every agent a chance (even if low) to interact or be interacted with at every time step. In the selective social influence model, only one agent is active at each time step and can only see messages from agents it concretely follows. However, messages can be seen later in time. Thus, connections in the selective social influence model are more formalized and are explicitly created and severed, whereas the activity-driven model's connections are better represented after a simulation by the number of interactions. Second, the peaks or echo chambers that form in the activity-driven model are symmetrical around 0, with only a maximum of two forming. The selective social influence model has asymmetrical peaks and can have up to six peaks\cite{sasahara2021social}. 
}
\mg{
Only the most robust RDN (Eq. \ref{RDNunique.eq}) was carefully evaluated in Fig. \ref{fig6.social_influence_model}, and therefore simpler terms in Table \ref{RDN.table} may not be applicable in alternative models. Furthermore, peak-promoting parameters (e.g. $\epsilon < 0.5$, "re-post" following strategy) may reduce or nullify the effects of the RDN. Nevertheless, the applicability to another model strengthens the case for the RDN to be a general principle of adding constructive noise to prevent or depolarize echo chambers.
}

Noise is a common factor that is always present when people make decisions \cite{kahneman2021noise}. The RDN introduced here could be considered as well-adjusted noisy input to the system which is determined by the collective dynamics of the network.  We argue that a well-adjusted noisy input such as a random dynamics nudge could be utilized to avoid the creation of extreme ideologies as well as echo chambers, at least in the case of opinion dynamics in social networks \cite{kahneman2021noise}.

It is important to note that recent research has shown that the presence of echo chambers might be overstated \cite{dubois2018echo,guess2021almost,eady2019many,bruns2017echo,barbera2015tweeting}, even though this appears to be conditional on the social media platform \cite{cinelli2020echo}. For example, \cite{guess2021almost} finds that echo chambers might not be an extended feature of social media platforms as previously thought and are limited to relatively few users. However, echo chambers have been observed and studied during events that concentrate the attention of users, political or otherwise \cite{cardenal2019echo,aruguete2018time}, and these events have the pernicious effects of echo chambers previously suggested, such as exacerbating polarization \cite{cinelli2021echo,du2016echo,balsamo2019inside}. While we do not debate the degree of prevalence of echo chambers in social media networks, the evidence points towards contexts that are ripe for the emergence of echo chambers and the consequences that accompany them.

As previously suggested, the formation of clusters in social media networks is a consequence of the selective exposure of users actively seeking cognitively congruent social media content \cite{himelboim2013tweeting, aruguete2018time}, high transitivity of social networks \cite{sasahara2021social,keijzer2018communication}, and reinforcing mechanisms advanced by the platform's algorithms \cite{cohen2018exploring}. Our RDN fits as an intervention within the latter (i.e., the platform's algorithms), part of the toolkit social media platforms possess to interact with their users. Social media platforms can identify events that polarize networks, as well as the polarized communities. Since the information used by our RDN comes from within the network (specifically, from those polarized communities), it is a plausible intervention mechanism that social media platforms can apply. Future extensions of the model aim to provide empirical evidence from the application of the intervention in actual online settings.

Possible theoretical extensions and feasible social network implementations of our model can focus on changes to different terms construing the RDN. For example, the RDN strength ($D$) plays a similar role to the interaction strength ($K$) but relates to a random sample of agents instead of the connections determined by agents' opinions. It is then reasonable that $D$ would naturally be a function of the issue at hand and the network's interaction strength. The content of RDN would also play a role and would be an integral part of determining adequate RDN strength. For example, one can show an aggregation of some users' opinions using numerical values of likes, retweets, engagement, etc. Relevant to RDN, adding randomness to each user's feed on a given topic may be a reasonable real-life approximation of the RDN. The frequency of these random posts, among other factors, could be a possible modulator of $D$. We note that understanding the nuances of user interface design approaches is out of this article's scope.

A limitation of this formulation compared to a social network is that the RDN is influential at every time point $dt$. It may be more realistic to consider a scenario where $D>0$ only if the agent is active. However, one can also consider the RDN as a background process of each agent that shifts their opinion even if they are not socially active. Giving an opinion to someone means that it will linger with them to reflect upon, even if they are not highly active on social media. To more accurately model this reflective behavior as a function of time, the RDN could be proportional to the agent's activity. Also, the results presented here are for a simple model, and a different picture might emerge when more complex interactions are considered amongst the agents. 

Additional work should be done to delineate the effects of choosing a random opinion, its continuous presentation, and its weighting. Focusing on the latter, extensions of the model can capture the expertise of an opinion and weight the exposure to random content based on this expertise. While our model tests a hypothetical scenario of polarization on a topic \textit{with no clear answer} and with equally-weighted agents, polarized opinions and echo chambers have been found to arise in discussions over topics that have an expert consensus (e.g., global warming). For these cases, randomly presenting weighted expert opinions (i.e., weighted on the distribution of opinions within the expert community) might be a more fruitful intervention. This approach would change the current approach to misinformation, which is to block content. More detailed models of opinion dynamics can make a unique contribution to understanding key challenges such as societal polarization \cite{flache2017models}. 

\section*{Conclusion}

In this paper, we used an opinion dynamics model that gives rise to echo chambers to study the effect of a random input from the network on itself. Motivated by the idea of the effect of nudges, we added a random dynamical nudge (RDN) as input from the network to each agent. Our results show that the RDN is an effective tool to stop the formation of echo chambers - or disband echo chambers already formed - by exposing agents to diverse views and counteracting the selective exposure of users. Thus, the RDN appears to be a robust and reliable method for addressing echo chambers.

Beyond the methodological contribution, the systems modeled in this paper reflect a polarized online environment that has become increasingly common. We believe that the RDN is a possible and viable solution that can encourage dialogue and, ultimately, strengthen democratic deliberation in social media. In the future, we hope to design real-life experiments either in-person or in online social networks to verify the findings of this paper.

\section*{code availability}
Full modeling details and the code are available online at \blue{\url{https://github.com/ChrisCurrin/opinion_dynamics}}.

\section*{Acknowledgements}

CBC and AKN would like to thank Neuromatch Academy \blue{\url{https://www.neuromatchacademy.org}}  for introducing the authors to each other. We thank Dr. Krešimir Josić (University of Houston) , Fabian Baumann (Humboldt University) and Dr. Igor M. Sokolov (Humboldt University) for carefully reading the early versions of the manuscript and providing constructive feedback. 

CBC is supported by the German Deutscher Akademischer Austauschdienst (DAAD, https://daad.de), the South African National Research Foundation (NRF, https://nrf.ac.za), the University of Cape Town (UCT, https://uct.ac.za), and the NOMIS Foundation through the NOMIS Fellowships at IST Austria program (https://nomisfoundation.ch).

\section*{Author contributions statement}
AKN defined the problem, CBC performed the numerical simulations, SVV helped to analyze the results, and all authors wrote and reviewed the manuscript. 

\bibliography{refs}

\end{document}